\documentclass{svjour3}                     
\smartqed  

\usepackage[numbers]{natbib}
\usepackage{graphicx}

\newcommand{\be}{\begin{equation}}
\newcommand{\ee}{\end{equation}}
\newcommand{\ba}{\begin{eqnarray}}
\newcommand{\ea}{\end{eqnarray}}
\newcommand{\bea}{\begin{eqnarray}}
\newcommand{\eea}{\end{eqnarray}}
\newcommand{\xx}{\bar{x}}          
\newcommand{\qq}{\bar{q}}          
\newcommand{\ggg}{\bar{g}}          
\newcommand{\nn}{\nonumber}

\begin{document}

\title{Vacuum gravitational fields with a null Killing vector}

\author{G. Dautcourt}
\institute{
Albert Einstein Institut, Golm, Germany \\
 \email{dautcourt@germanynet.de}   }

\date{Received: date / Accepted: date}

\maketitle 

\begin{abstract}
Vacuum gravitational fields admitting a light-like Killing field were systematically studied starting around 1960. Besides the already known plane waves, a second class of gravitational wave fields was found. In contrast to plane waves, their wave surfaces were not flat, but had a negative Gaussian curvature. Recently, such solutions found attention again as ``twisted gravitational waves". In the paper we review and extend the earlier results. In suitable coordinates, the metric assumes a simple shape. The waves are then determined by a single function that satisfies a Laplace equation in cylindrical coordinates. The ``twisted waves" prove to be a special case.                            
\end{abstract}

\maketitle

\section{History and summary}

Einstein hoped to explain the discrete quantum structure of matter in terms of singularity-free solutions of a nonlinear unified field theory \cite{Ein}. In particular, the graviton was expected to be represented by an exact solution of the nonlinear vacuum field equations of gravity.
What would this solution look like? Analogous to a photon, the graviton had to be a massless particle moving with the velocity of light. Also, the field configuration of a single graviton without interaction was expected to be preserved during its motion in a direction $k^\mu$. These two conditions led to specific geometric requirements: $k^\mu$ had to be a null vector in the geometry created by the graviton, and this geometry should not change as the graviton moves along the integral curves of $k^\mu$. 
This means, that \emph{the single classical graviton in Einstein's vision should admit a null Killing field}, just as an isolated particle with non-zero rest-mass has a time-independent gravitational field, admitting a time-like Killing vector.
           
Such heuristic ideas were discussed 
at the beginning of the Sixties of the last century
in Papapetrou's research group at the former Prussian Academy of Sciences in Berlin. 
The author as a member of the group worked on the characteristic initial value problem for the vacuum field equations of General Relativity at that time \cite{Daut1,Daut2}. 
His method was not applicable in the case that the tangent vector of the initial null hypersurface was a Killing vector (that is, if the hypersurface was a horizon). This special case had to be treated separately\footnote{see the note added in proof of \cite{Sachs62}, p. 914.}.    
                             
So there was ample reason to handle vacuum fields with a null Killing vector. 
Surprisingly, it was pretty easy to find the general solution \cite{Daut1,Daut3}. 
There were two types of solutions, depending on the Killing field being a gradient or not. The first case corresponded to the - already at that time - well known plane waves. In the second case a new class of exact solutions emerged, also with apparently free functions.
The essential difference between both types is the character of their wave surfaces, two-dimensional surfaces orthogonal to the propagation direction of the wave. 
For a gradient Killing field the wave surfaces are Euclidean planes with zero Gaussian curvature - hence the name ``plane waves"\footnote{We consistently use the term ``plane wave" for what is often referred to as ``pp wave".}.   
 The wave surfaces  of the new class had non-constant negative Gaussian curvature.

Plane waves have little similarity with Einstein's classical graviton problem, but also the new class of solutions was a disappointment in this respect. 
Just like many other solutions to Einstein's vacuum equations, the new metrics had true singularities
that could not be removed by a coordinate transformation.
Freely specifiable functions could be interpreted as wave amplitudes, and the solutions
led asymptotically to plane waves - features which could qualify the metric as a gravitational wave -, but the singularities were hard to explain. Thus only a short communication was published at the time. 
The fields received little attention until a recent rebirth as ``twisted gravitational waves"  \cite{Bini1}-\cite{Fir}.   
                                                
As explained in section 2, all these solutions belong to Kundt's class of vacuum fields. 
In section 3 and 4 we derive the general solution.
We closely follow the original treatment, which is otherwise only available in German.
In section 5 an interpretation as a gravitational wave is attempted and 
some special cases are discussed.     
For the sake of completeness, section 6 shortly deals with plane waves, whose interpretation as gravitational waves is undisputed. Section 7 explains in some detail the relation to the ``twisted gravitational waves" (TGW). It is shown that the TGW's are special cases of our solution.

The article deals exclusively with vacuum fields. At the end we 
briefly consider the extension to non-vacuum gravitational fields.

\section{Relation with Kundt's class}

A null Killing field $k^{\mu}$ satisfies the relations\footnote{Greek letters range from 0 to 3, Latin letters from 1 to 3, Capital latin letters from 2 to 3. Other conventions are those of \cite{R1} or \cite{Wald}.}
\bea \label{ke}
k_{\mu ; \nu} + k_{\nu ; \mu} &=& 0, \\
k^\mu k^\nu g_{\mu\nu}  &=& 0. \label{norm} 
\eea
It follows $k_{\mu;\nu}k^\nu = 0,$ so the field $k^\mu$ represents a geodesic null congruence.
A general geodesic null congruence $k^\mu$ can be characterized by several scalars, obtained by decomposing the covariant derivative of $k_\mu$. Expansion $\theta$, the square of the shear amount $|\sigma|^2$ and twist $\omega$ are given by
\bea 
\theta &=& \frac{1}{2}k^\mu_{;\mu}, \\
|\sigma|^2 &=& \frac{1}{2}k_{(\mu;\nu)}k^{\mu;\nu} - \frac{1}{4}(k^\mu_{;\mu})^2,\\
\omega^2 &=& \frac{1}{2}k_{[\mu;\nu]}k^{\mu;\nu}.
\eea
The scalars are related by the Raychaudhuri equation
\be \label{rel}
\theta_{,\mu}k^\mu - \omega^2 + \theta^2 + |\sigma|^2 = -\frac{1}{2}R_{\mu \nu}k^\mu k^\nu.
\ee
Obviously, the Killing congruence defined by (\ref{ke}) and (\ref{norm}) is a non-expanding shear-free geodesic null congruence, but not apriori twist-free. (\ref{rel}) shows that for vacuum fields also the twist vanishes.

Kundt has studied a number of gravitational fields related to the problem \cite{Kundt61,EK,KuTr}.
His class of solutions assumes a geodesic null congruence with zero expansion and twist. For vacuum fields one concludes from (\ref{rel}) that Kundt's fields are also shear-free. 
 Thus our solution belongs to Kundt's class.                                     

\section{The field equation $R_{\mu\nu}k^\mu k^\nu =0$ in adapted coordinates}

For the integration of the field equations it is useful to introduce coordinates adapted to the symmetry. We choose $k^\mu$ as tangent to the $x_0$-lines:
\be
k^\mu = \delta^\mu_0,
\ee
thus (\ref{ke}),(\ref{norm}) reduce to 
\be \label{cond1}
g_{\mu\nu , 0} = 0,~~  g_{00}= 0.
\ee
The second condition $g_{00}=0$ is equivalent to $|g^{ik}|=0$. Since the four-dimensional metric is nonsingular by assumption, $|g^{\mu\nu}|\neq 0$,  we have $rank(g^{ik})=2$. Thus there exists at least one two-dimensional submatrix of $(g^{ik})$ with non-vanishing determinant, say
\be
|g^{AB}| = g^{22}g^{33}-g^{23}g^{23} \neq 0.
\ee
With the quantities
\be
\mu =  (g^{12}g^{33}-g^{13}g^{23})/|g^{AB}|, ~ 
\lambda = (g^{13}g^{22}-g^{12}g^{23})/|g^{AB}|   
\ee
we can write
\be
|g^{\mu\nu}| = - |g^{AB}|(g^{00} -\mu g^{02} - \lambda g^{03})^2.
\ee
This implies $|g^{AB}| > 0$.

We try to simplify the metric by transforming $x_1$ into a null coordinate. One of the coordinate transformations preserving the conditions (\ref{cond1}) is 
\be \label{t1}
\xx_0   = x_0, ~~ \xx_1   = f(x_1, x_A), ~~ \xx_A = x_A  
\ee
with a function $f(x_1,x_A)$.  The condition $\ggg^{11} =0$ for a null coordinate $\xx_1$ takes the form                           
\be \label{si}
\sigma_A\sigma_B g^{AB} = 0
\ee
with 
\be
\sigma_2 = \mu f_{,1} + f_{,2}, ~~\sigma_3 = \lambda f_{,1} + f_{,3}.
\ee
Since $|g^{AB}| >0$, one concludes $\sigma_A =0$ or explicitly
\be
\mu f_{,1} + f_{,2} = \lambda f_{,1} + f_{,3} = 0.
\ee
This system of differential equations for the single function $f$ requires 
\be
 \mu_{,3} - \lambda_{,2} + \lambda \mu_{,1} - \mu \lambda_{,1} = 0
\ee
as integrability condition. Surprisingly, the field equation
 $R_{\mu\nu}k^\mu k^\nu =0$ or in adapted coordinates
\be \label{r00}
R_{00} = -\frac{1}{2} g_{01}^2(\mu_{,3} -\lambda_{,2} + \lambda \mu_{,1} - \mu \lambda_{,1})^2 = 0  
\ee
provides just this integrability condition. Moreover, the transformation (\ref{t1}) leads also to
\be
\ggg^{1A} = g^{1A}f_{,1} + g^{AB}f_{,B} = 0.
\ee
 Finally, we may reach for the two-dimensional metric $g_{AB}$ a diagonal form $g_{AB} = e^{\phi}\delta_{AB}$ by means of 
\be \label{tr0}
\xx_0= x_0,~~ \xx_1 = x_1,~~\xx_A = x_A(x_i).
\ee
After these simplifications, the line element becomes (omitting the bar)               
\be \label{le}
ds^2 = 2g_{01}dx_0dx_1 + g_{11} dx_1^2 + 2g_{1A}dx_1dx_A + e^{\phi}( dx_2^2 + dx_3^2).
\ee  
The five field functions $g_{1i}$ and $\phi$ depend on $x_i$. This form of the line element is preserved under the transformations                 
\be \label{tr}
\xx_0= x_0 + \varphi(x_i),~~\xx_1 = \xx_1(x_1),~~\xx_A =  \xx_A(x_i) 
\ee 
with  arbitrary functions $\varphi(x_i),~\xx_1(x_1)$, only $\xx_A(x_1,x_A)$ has to satisfy 
the two-dimensional Laplace equation 
\be
\Delta \xx_A \equiv  \frac{\partial^2 \xx_A}{(\partial x_2)^2 } + \frac{\partial^2 \xx_A}{(\partial x_3)^2 } =0. \label{lap} 
\ee
The metric (\ref{le}) already satisfies the vacuum equations $R_{00}=0$ and $R_{0A}=0$. One can use  
coordinate transformations still available to simplify and solve the other field equations. 

\section{Solving the remaining field equations}

Some remaining vacuum equations, written as $ R_{22}- R_{33}, R_{23}, R_{01}, R_{22}+ R_{33}$,  are respectively
\bea
 \phi_{,2} m_{,2} -\phi_{,3} m_{,3} = m_{,22} - m_{33} +(m_{,2})^2/2 -(m_{,3})^2/2 \equiv   A,  \label{r+}  \\ 
 \phi_{,2} m_{,3} +\phi_{,3} m_{,2} =  m_{,2} m_{3} +2 m_{,23} \equiv   B,  \label{r-}\\
 \Delta m +(m_{,2})^2 + (m_{,3})^2 = 0, \label{r01} \\
  \Delta\phi   + \Delta m + (m_{,2})^2/2 +(m_{,3})^2/2=0. \label{r23}
\eea
Here we have set $g_{01}= e^m$ (we can assume that $g_{01}$ has the same sign everywhere, a zero would mean $|g_{\mu\nu}| =0$).
Under the assumption $(m_{,2})^2 + (m_{,3})^2 \neq 0$ 
we solve (\ref{r+},\ref{r-}) for $\phi_{,2}$ and 
 $\phi_{,3}$:
\bea
\phi_{,2} & = & (A m_{,2} +B m_{,3})/((m_{,2})^2 + (m_{,3})^2),\\
\phi_{,3} & = & (B m_{,2} -A m_{,3})/((m_{,2})^2 + (m_{,3})^2).
\eea
A short calculation shows, that because of (\ref{r01}), the integrability condition $\phi_{,23} =\phi_{,32}$ as well as the field equation (\ref{r23}) are already satisfied.

The coordinates $x_A$ are only fixed up to harmonic transformations with (\ref{lap}).
 Following Weyl \cite{Weyl}, we introduce special canonical coordinates. Let $\xx_2 = f(x_1,x_2,x_3)$ be a set of functions of $x_A$, parametrized by $x_1$, which satisfies $\Delta f =0$.
Then there always exists another set of harmonic functions, say $\xx_3= g(x_1,x_2,x_3)$ , satisfying the Cauchy-Riemann equations $f_{,2}=g_{,3}, ~ f_{,3}= -g_{,3}$. Since (\ref{r01}) can be written $\Delta e^m = 0$, 
$e^m$ is already such a harmonic function. Thus we are able to set $e^m= x_2$ 
(we can imagine $x_2$ as a kind of radial coordinate).
 The determination of $\phi$ from (\ref{r+},\ref{r-}) now becomes trivial and leads to 
\be
e^{\phi} = a(x_1)/\sqrt{x_2},
\ee
where $a(x_1)$ is a free function.

 Introducing canonical coordinates $x_A$ also simplifies the integration of the remaining field equations $R_{1i}=0$. Writing $g_{12} = x_2 q_2, ~g_{13}= x_2 q_3$ with two functions $q_A(x_1,x_2,x_3)$, the conditions $R_{12}=0$ and $R_{13}=0$ lead to
\bea
q_{3,23} -q_{2,33} + a_1/(x_2)^{5/2}= 0, \\
q_{2,32} - q_{3,22} + 5 (q_{2,3} - q_{3,2})/x_2 = 0,              
\eea
which can be integrated to give
\be \label{q}
q_{2,3}-q_{3,2} = (a_{,1}x_3 - b)/(x_2)^{5/2}.
\ee
$b=b(x_1)$ is a second free function. We still have one transformation $\xx_0 = x_0 + \varphi(x_i),~ \xx_i=x_i$  at our disposal, which leads to 
\be \label{tr1}
\qq_A =  q_A + \varphi_{,A} 
\ee
and will be used to reach $q_2=0$. Then from (\ref{q}) after integration 
\be
q_3 =  \frac{2}{3}(a_{,1}x_3 - b)/(x_2)^{3/2}.                
\ee
Here a function $c(x_1,x_3)$ resulting from this integration has been transformed to zero 
with help of (\ref{tr1}). The last equation $R_{11}=0$ can now be written\footnote{The corresponding equation (28) in \cite{Daut3} is misprinted: instead of 4/3 read 2/3. The misprint  was noted in \cite{Daut4}. Also equation (30) in \cite{Daut3} is misprinted: instead of $3 a^2/(4x_2^3)$ read $3/(4 a^2 x_2^3)$.}    
\be \label{feq}
\frac{\partial^2H}{\partial x_2^2} + \frac{1}{x_2} \frac{\partial H}{\partial x_2}  + \frac{\partial^2H}{\partial x_3^2}  =  x_2^{-7/2} (b-a_{,1}x_3)^2/a + x_2^{-3/2}(a_{,1}^2/a -2 a_{,11}/3), 
\ee
where we have set $g_{11}= rH(x_1,x_2,x_3)$.
The general solution  $H$ of (\ref{feq}) is represented as the sum of a particular solution and the general solution of the homogeneous equation
\be \label{hom}
\frac{\partial^2F}{\partial x_2^2} + \frac{1}{x_2} \frac{\partial F}{\partial x_2}  + \frac{\partial^2F}{\partial x_3^2}  =  0.
\ee
It is not difficult to find a particular solution of the inhomogeneous equation, so the general solution of (\ref{feq}) can be written 
\be \label{ffull}
H = F +\frac{4}{9}x_2^{-3/2}(b-a_{,1}x_3)^2/a + x_2^{1/2}(\frac{4}{9}a_{,1}^2/a -\frac{8}{3} a_{,11} ).  \ee
The metric now becomes
\be \label{me}
ds^2 = 2x_2dx_0dx_1 + Hx_2 dx_1^2 +\frac{4}{3\sqrt{x_2}}(a_{,1}x_3 -b) dx_1dx_3 + \frac{a}{\sqrt{x_2}}(dx_2^2 +dx_3^2),
\ee  
with two arbitrary functions $a(x_1)$ and $b(x_1)$. In this form the metric was published \cite{Daut3}.
It is not mentioned in \cite{Daut3} that the functions $a(x_1)$ and $b(x_1)$ are still subject to coordinate changes of the type (\ref{tr}) and (\ref{lap}). The transformations                                             
\be
\xx_0 = x_0 +\varphi, ~ \xx_1 = q(x_1), ~ \xx_2 = x_2/q_{,1}, ~ \xx_3 = p(x_1) +x_3/q_{,1}     
\ee
with two arbitrary functions $p(x_1)$ and $q(x_1)$  ($q_{,1} \neq 0$) send $a$ und $b$ to new functions
\be 
a \rightarrow a (q_{,1})^{3/2}, ~ b \rightarrow (2b +2a_{,1}p q_{,1} +3 a p_{,1}q_{,1} + 3apq_{,11} )/(2 \sqrt{q_{,1}}).
\ee   
In particular, choosing $p$ and $q$ according to 
\be
q_{,1} = a^{-2/3},~ p_{,1} = -\frac{2}{3} ba^{-1/3}   
\ee
we may reach standard values $a=1$ and $b=0$. With this gauge the metric becomes diagonal, depending only on the function $F(x_1,x_2,x_3)$: 
\be  \label{stan}
ds^2 = 2x_2 dx_0dx_1 + x_2F dx_1^2 + \frac{1}{\sqrt{x_2}}(dx_2^2 + dx_3^2).
\ee
For later use we note another form of the metric with an arbitrary non-vanishing spurious function $\tau(x_1)$, allowing for self-similar changes of $x_1$, also a constant $a \neq 1$ is left:
\be \label{mets}
ds^2 = 2 \tau x_2 dx_0dx_1 + F\tau^2 x_2 dx_1^2 +  \frac{a}{\sqrt{x_2}}(dx_2^2 +dx_3^2).
\ee

In deriving (\ref{stan}) starting from the definition equations (\ref{ke}),(\ref{norm})
  we have made no further restricting assumptions, except of $(m_{,2})^2 + (m_{,3})^2 \neq 0$ (dropping this assumption leads to the plane wave branch). We can therefore expect that (\ref{stan}) represents the general solution to the problem under discussion.

Solutions of the vacuum equations admitting non-isotropic Killing vectors exist in large numbers \cite{R1}. It is perhaps noteworthy that \emph{in the case of a null Killing field
 (with $k_{\mu;\nu} \neq 0$) only one simple solution class (\ref{stan}) exists.  
The solution is governed by a single function $F(x_1,x_2,x_3)$, which satisfies the linear differential equation (\ref{hom})}. For the other solution class with $k_{\mu;\nu} = 0$, the plane waves, a similar statement applies (section 6).                                             

\section{Properties of the solutions}
The null hypersurfaces $x_1=const$ with $k^\mu$ as tangential vector can be regarded as propagation fronts of a gravitational wave. Each three-dimensional propagation front is formed by a set of two-dimensional wave surfaces orthogonal to $k^\mu$. Wave surfaces can be used to define an amplitude of the wave:
Besides $k^\mu$, there is another null direction orthogonal to a wave surface, say $l^\mu$. 
Geodesic continuation of $l^\mu$ generates a null hypersurface for every wave surface, which is called \emph{conjugate} to the propagation front $x_1=const$. Then part of the geometry of the conjugate hypersurfaces at their intersection with the wave surfaces can serve as \emph{geometrical measure of the wave intensity}. 
The required data are those which could also be taken as initial data in a characteristic initial value problem related to the null hypersurfaces. A convenient datum is the complex Penrose function
\be
            P = |P| e^{i\theta}= R_{\mu\nu\rho\sigma} l^\mu\bar{m}^\nu l^\rho\bar{m}^\sigma
\ee
with amplitude $|P|$ and phase $\theta$.  $m^\mu$ is a complex null vector spanning the directions in the wave surfaces, $\bar{m}^\mu$ complex-conjugated. The Penrose function depends only on the \emph{inner} geometry of the conjugated null hypersurfaces. 
For the general metric (\ref{stan})  one obtains
\be \label{Pen}
 P = -\frac{1}{2}x_2^{-3/2}(x_2 \frac{\partial^2 F}{\partial x_2^2} + \frac{5}{4} \frac{\partial F}{\partial x_2}) 
+ \frac{i}{2} x_2^{-3/2}(x_2 \frac{\partial^2 F}{\partial x_2 \partial x_3} + \frac{3}{4} \frac{\partial F}{\partial x_3}). 
\ee

This simple interpretation of the metric as a progressing gravitational wave cannot hide that one is far from a physical understanding.  
The outstanding feature of the solution is the singularity for $x_2 \rightarrow 0$, 
for which no convincing physical interpretation seems to be available. 
The invariants of the Riemann tensor
\bea
I_1 &=& R_{\mu \nu \rho \sigma}\,R^{\mu \nu \rho \sigma} - i R_{\mu \nu \rho \sigma}\,R^{*\,\mu \nu \rho \sigma}\ = \frac{3}{4}x_2^{-3} , \\
I_2 &=& R_{\mu \nu \rho \sigma}\,R^{\rho \sigma \alpha \beta}\,R_{\alpha \beta}{}^{\mu \nu} + i R_{\mu \nu \rho \sigma}\,R^{\rho \sigma \alpha \beta}\,R^{*}{}_{\alpha \beta}{}^{\mu \nu}\ 
= -\frac{3}{16}x_2^{-9/2}
\eea                  
show this singularity. The singularity is also visible in the Gaussian curvature of the wave surfaces:
\be \label{gc}
K = -\frac{1}{4}x_2^{-3/2}.
\ee
For large $x_2$ the wave surfaces become flat and the invariants tend to zero. These are signs that the fields  
asymptotically tend to plane waves. 

We leave the problem of singularities open for the time being, and
 move on to some explicit solutions.
(\ref{hom}) has the form of a three-dimensional Laplace equation in cylindrical coordinates $(r=x_2, z=x_3)$ for a function $F$ with axial symmetry, i.e. independent of the azimuth.
The trivial case $F=0$ leads to 
\be \label{f0}
ds^2 = 2x_2 dx_0dx_1  + \frac{1}{\sqrt{x_2}}(dx_2^2 + dx_3^2).
\ee
This is one of the simplest curved Ricci-flat geometries.
The metric can be transformed into one of the Kasner solutions \cite{Kasner} and
was discovered independently several times.                          
Harrison found the solution as case III-2 among his 30 Ricci-flat metrics \cite{Harr}, but in coordinates which hide its simplicity. The solution also belongs to the so-called ``one-dimensional gravitational fields" discussed in \cite{Daut0}.

 (\ref{f0}) admits two independent null Killing vectors 
$k^\mu =\delta^\mu_0,~ l^\mu= \delta^\mu_1$  and is the only metric in our class with this property.
The Petrov type is D.                                              
The two sets of null hypersurfaces $x_0 = const$ and $x_1=const$ are both horizons with vanishing shear and expansion of their inner geometries. Their intersections are two-dimensional surfaces with the negative Gaussian curvature (\ref{gc}).                     
 (\ref{f0}) can be considered as the unique solution of a characteristic initial value problem starting from a pair of conjugated horizons together with a given two-dimensional geometry of their intersection. 
If one writes the metric in space-time coordinates, its static character becomes visible, so it is certainly not a gravitational wave. A wave intensity formally calculated with (\ref{Pen}) is zero. The solution is however the key to a physical understanding of equally singular solutions, that clearly show a wave character.

A function $F=f(x_1)$, also a trivial solution of (\ref{hom}), can be reduced to $F=0$ by a coordinate transformation. Simple, but non-trivial solutions result from a separation ansatz $F=A(x_1,x_2) B(x_1,x_3)$,  leading to
\be \label{sep}
\frac{1}{A}\frac{d^2A}{dx_2^2} + \frac{1}{x_2}\frac{1}{A}\frac{dA}{dx_2}= -k,
	~ \frac{1}{B}\frac{d^2B}{dx_3^2} = k,                  
\ee
$k$ may depend on $x_1$.  For $k=0$, this effectively gives                       
\be
 F =U(x_1)x_3 + V(x_1) ln x_2 + W(x_1) x_3 ln x_2.
\ee
We interpret this metric as superposition of three gravitational waves, which transmit information stored in the functions $U$, $V$ and $W$ of $x_1$. The total amplitude and phase as measured by the Penrose function are
\be
P = \frac{3i}{8} x_2^{-3/2} U - \frac{1}{8}x_2^{-5/2}V -\frac{1}{8}x_2^{-5/2}x_3 W + \frac{i}{2} x_2^{-3/2}(1+ \frac{3}{4} ln x_2)W.
\ee
From the Penrose function we can read off the different polarization modes of the $U,V$ and $W$-waves: The real part of $P$ corresponds to $\oplus$-polarization, the imaginary part to $\otimes$-polarization. It is seen that the $U$-wave has $\otimes$-polarization, the $V$-wave $\oplus$-polarization. $W$-waves show both types of polarization.

If in (\ref{sep}) $k$ differs from zero and is positive, one obtains solutions involving Bessel functions. A basic solution with Bessel functions of the first kind is  
\be \label{bes1}
 F= J_0(\sqrt{k}x_2)e^{\pm \sqrt{k}x_3}.
\ee
The corresponding polarization amplitudes are:  
\bea \label{bes2}
P_{\oplus}  &=& \frac{e^{\pm \sqrt{k}x_3}}{8x_2^{3/2}}
 ( \sqrt{k} J_1(\sqrt{k} x_2) +4kx_2J_0(\sqrt{k}x_2) ),    \\
P_{\otimes} &=& \frac{e^{\pm \sqrt{k}x_3}}{8x_2^{3/2}} 
 ( \mp 4 J_1(\sqrt{k} x_2) \pm 3 \sqrt{k} J_0(\sqrt{k}x_2)).\label{bes3}           
\eea
Another solution results, if the Bessel functions of the first kind in (\ref{bes1})-(\ref{bes3}) are replaced by Bessel functions of the second kind (Weber functions), with different behaviour at the singularity $x_2 \rightarrow 0$ and for $x_2 \rightarrow \infty$.

For a negative $k$ one has similar expressions with waves oscillating in $x_3$. 
Because of the linearity of (\ref{hom}), all solutions can be superimposed linearly with coefficients depending arbitrarily on $x_1$.

One may think of other solutions, but the given examples are sufficient to illustrate the class. The Petrov type is in general II, in special cases D. 

\section{Plane waves} 
To complete the discussion, we shortly consider plane waves, the other class of vacuum solutions in case of a null Killing vector. 
There were many papers on plane waves around 1960 (and earlier), either with  Brinkmann-like coordinates or in the Rosen form  
\cite{Brinkmann,BJ,Ros,Tak58,BPR59,Peres59,Tak60,Kundt61,EK,KuTr}. 
Plane waves result if we drop the condition
$(m_{,2})^2 + (m_{,3})^2 \neq 0$ and assume $m=m(x_1)$. It is easy to see that the Killing vector is a gradient (and therefore covariantly constant), if and only if $m$ is independent of $x^A$. 
Again we can use coordinate transformations to simplify the solution.
For Brinkmann coordinates, we can write the line element (for a derivation, see e.g. \cite{R1})
\be \label{pp1}
ds^2 = 2dx_0dx_1 + A dx^2_1 +  dx_2^2 + dx_3^2.
\ee
As in our previous case, a single function $A=A(x_1,x_2,x_3)$ governs the solutions, but now $A$ satisfies (in the vacuum case) the two-dimensional Laplace equation $\Delta A =0$. We take the local solution which is quadratic in the transversal coordinates $x^A$:
\be
A = \alpha(x_1)(x^2_3-x^2_2) + 2 \beta(x_1)x_2 x_3.
\ee
Also here the null hypersurfaces $x_1=const$ serve as propagation fronts. The complex Penrose function of the conjugated null hypersurfaces as a measure of the wave intensity reduces to 
\be
      P = \alpha(x_1)  + i \beta(x_1), 
\ee
in accordance with our expectation. The wave surfaces are Euclidean planes.

\section{``Twisted gravitational waves"}
Recently solutions admitting a null Killing field were considered in a series of papers \cite{Bini1,Bini2,Rosq,Fir} under the heading ``twisted gravitational waves" (TGW).
The authors have carried out an extensive study of some of these metrics, including an investigation of particle movements. 

 The term ``twisted" is used here in an unusual way: 
In the standard literature \cite{EK,R1,Griff,Wald}  a congruence $k^\mu$ is called twisted, if  
$k_{[\mu;\nu}k_{\rho]} \neq  0 $, 
which is equivalent to the non-vanishing of the twist-scalar  $\omega$.
Instead, the authors call solutions twisted, if $k_{[\mu;\nu]}$ differs from zero. Thus ``twisted metrics" in this unconventional sense may have (and in case of the above papers actually do have) a vanishing twist-scalar $\omega$.

The authors consider solutions $\Psi(\xx_1, \xx_2)$ of the partial differential equation 
\be \label{S1}
\frac{\partial}{\partial \xx_2}\left(\Psi \frac{\partial^2 \Psi}{\partial \xx_1^2}\right) =0, 
\ee
also written  
\be \label{S2}
\Psi \frac{\partial^2\Psi}{\partial\xx_1^2} = V(\xx_1) 
\ee    
with an  arbitrary function $V(\xx_1)$. 
Then the metric given by 
\be \label{S0}
ds^2 =  -\Psi^4 d\xx_0d\xx_1 + \lambda^2\,\Psi^4\,\left(\frac{\partial\Psi} {\partial \xx_2}\right)^2  d\xx_2^2 + \frac{1}{\Psi^2}\, d\xx_3^2
\ee
($\lambda > 0$ is a constant) is  Ricci-flat and admits a null Killing vector $k^\mu = \delta^\mu _0$. The metric is only implicitly known, since it depends on a solution of the differential equation (\ref{S1}). 

Since we claim to have found the general Ricci-flat solution in the null Killing vector case, given by (\ref{stan}) or (\ref{mets}), there must exist a coordinate transformation that relates a solution of the form (\ref{S0}) to our solution. 

TGW coordinates are noted as $\xx_\mu$, our coordinates are $x_\mu$. The null Killing vector has in both metrics the components $k^\mu = \delta^\mu_0$. Coordinate transformations preserving this property are again of the type (\ref{tr}). Rewritten in inverse form we have
\bea
x_0  & = & \xx_0 - \varphi(\xx_i), ~ x_1  = \xx_1,~\nn \\ 
x_2  & = & h(\xx_i),~ x_3  =  k(\xx_i). 
\eea
Using (\ref{mets}) with the spurious function $\tau(x_1)$ instead of (\ref{me}) 
allows to use the same coordinate $x_1$ in both metrics.

We first consider some explicit solutions discussed in \cite{Bini1}-\cite{Rosq}.
One of them 
is the Harrison metric \cite{Harr}, given by (using the form found in \cite{Rosq})
\be
ds^2 = -\xx_2^{4/3}d\xx_0d\xx_1 + \xx_1^{6/5}d\xx_2^2 + \xx_2^{-2/3}\xx_1^{-2/5}d\xx_3^2.
\ee 
In these coordinates the Harrison metric does not have the standard structure (\ref{S0}), but still belongs to the TGW class. The transformation
\bea
x_0 &=& \xx_0 + \frac{9}{5} \xx_1^{1/5} \xx_2^{2/3},~  x_1 = \xx_1, \nn  \\ 
x_2 &=& (3/4)^{4/3} \xx_1^{4/5} \xx_2^{4/3}, ~ x_3 = (3/4)^{1/3} \xx_3    
\eea
leads from Harrison to the solution (\ref{mets}) with
\be
a = 1,~ F=0, ~ \tau= -\frac{2^{5/3}}{3^{4/3}} x_1^{-4/5}.
\ee

Another simple example is the so-called $w$-metric, arising from (\ref{S0}) with $\Psi=\xx_1+\xx_2$ and $\lambda =1$:
\be
ds^2 = -(\xx_1+\xx_2)^4 d\xx_0d\xx_1 + (\xx_1+\xx_2)^4 d\xx_2^2 +(\xx_1+\xx_2)^{-2} d\xx_3^2.   
\ee
Here the transformations
\bea
x_0 &=& \xx_0 +\xx_1  + 2\xx_2, ~  x_1 = \xx_1,  \nn \\
x_2 &=& (\xx_1 + \xx_2)^4/2, ~ x_3 = 2 \xx_3 
\eea
generate (\ref{mets}) with 
\be
a = \frac{1}{4\sqrt{2}}, ~F=0, ~ \tau= -1.  
\ee
Thus the transformed $w$-metric essentially coincides with the transformed Harrison metric: 
Obviously, ``Harrison" and ``$w"$ represent the same metric in different coordinate systems, both of which are equivalent to the basic solution (\ref{f0}).
The relations                   
\bea
\hat{x}_0 &=& \xx_0 + 9\xx_1^{1/5}\xx_2^{2/3}/5-2\xx_1^{1/5}\xx_2^{1/3}/3^{-1/3} + 5\xx_1^{1/5}/3^{4/3}, \nn \\
\hat{x}_1 &=& 5 \xx_1^{1/5}/3^{4/3},~ \hat{x}_2 = 3^{1/3}\xx_1^{1/5}\xx_2^{1/3} - 5 \xx_1^{1/5}/3^{4/3},~ \hat{x}_3  = 3^{1/3}\xx_3
\eea
give the $w$-metric coordinates $\hat{x}_\mu$ in terms of the 
Harrison coordinates $\xx_\mu$.

Turning now to the general case (\ref{S0}), we show that the relations
\bea
x_0 &=& \xx_0 - \varphi(\xx_1,\xx_2,\xx_3),  \nn \\
x_1 &=& \xx_1, ~x_2 = h(\xx_1,\xx_2), ~ x_3 = 2 \xx_3/\lambda  \label{tr4}
\eea
with
\be \label{h}
h(\xx_1,\xx_2) = \Psi^4(\xx_1,\xx_2)/2
\ee
and suitable $\varphi$ transform the TGW metric into (\ref{mets}) with $\tau =-1$.                     
The transformation equations 
\[  g^{\mu\nu} = \frac{\partial x_\mu}{\partial \xx_\alpha} \frac{\partial x_\nu}{\partial \xx_\beta} \bar{g}^{\alpha\beta} \] 
give for the index pair $(\mu,\nu)$=(2,2) or (3,3)
\be
 a = \frac{\lambda^2}{4\sqrt{2}}.
\ee
The pairs (0,0), (0,1) and (0,3) provide the derivatives $\varphi_i= \frac{\partial \varphi}{\partial\xx_i} $. The remaining equations are already satisfied. We find 
\be 
\varphi_1  =  -\lambda^2\left(\frac{\partial \Psi}{\partial \xx_1}\right)^2 -\frac{\bar{F}}{2}, ~ 
\varphi_2 = -2\lambda^2\frac{\partial \Psi}{\partial \xx_1}\frac{\partial \Psi}{\partial \xx_2}, ~
\varphi_3 = 0,                             
\ee 
where the constancy of $a$ is already taken into account.
 The  integrability conditions require $\varphi_{2,3}- \varphi_{3,2}=0, ~\varphi_{3,1}-\varphi_{1,3}=0, ~\varphi_{1,2}- \varphi_{2,1}=0.$       
The last two give        
\be \label{res}
\frac{\partial \bar{F}}{\partial \xx_3} =0, 
~ \frac{\partial \bar{F}}{\partial \xx_2}= 
4 \lambda^2 \frac{\partial^2\Psi}{\partial \xx_1^2} \frac{\partial\Psi}{\partial \xx_2 }.
\ee
The function $F(x_1,x_2,x_3)$ is written here $\bar{F}(\xx_1,\xx_2,\xx_3)$ as function of the TGW coordinates:
\be
\bar{F}(\xx_1,\xx_2,\xx_3)\equiv F(\xx_1,x_2[\xx_1,\xx_2], x_3[\xx_1,\xx_3]).
\ee
In original coordinates $F$ satisfies the linear equation (\ref{hom}). Translated into TGW coordinates by means of (\ref{tr4}), the transformed function $\bar{F}$ satisfies 
\be \label{homtr}
\frac{\partial^2\bar{F}}{\partial \xx_2^2}+ \left(\frac{h_2}{h} - \frac{h_{22}}{h_2}\right)\frac{\partial \bar{F}}{\partial \xx_2} + \frac{\lambda^2 h^2_2}{4}\frac{\partial^2\bar{F}}{\partial \xx_3^2} = 0,
\ee
where $h_2= \frac{\partial h}{\partial \xx_2}$ etc. Inserting (\ref{h}) and (\ref{res}) we find
\be \label{homtr1}
\frac{\partial \Psi}{\partial \xx_2}\left(\Psi \frac{\partial^3 \Psi}{\partial \xx_2 \partial \xx_1^2} 
+ \frac{\partial^2\Psi}{\partial \xx_1^2} \frac{\partial \Psi}{\partial \xx_2}\right) =0.
\ee 
Since $\frac{\partial \Psi}{\partial \xx_2} \neq 0$, the basic equation (\ref{S1}) for $\Psi$ is recovered.
Clearly, this relation must be satisfied, thus that metrics of the type (\ref{S0}) can be transformed into our metric. For the function $\bar{F}$ we obtain 
\be
 \bar{F} = 4\lambda^2 V(\xx_1) \ln{\Psi} 
\ee
up to an added arbitrary function of $\xx_1$. Translated 
into our coordinates we have the result:
\emph{The TGW metrics are the special case $F= V(x_1) \ln x_2$ of our solution. In particular, the Harrison metric (or w-metric) corresponds to $F=0$}.

\section{Final remarks}

Many of the issues concerning nonplanar waves need further clarification. In recent years the ``cosmic jet" property was intensively investigated: Test particles can gain (or lose) energy in time-dependent gravitational fields,  e.g. plane waves \cite{Bini01,Bini02}. 
It would be interesting to study this question also for the nonplanar waves presented in this article. 

Another task is the extension beyond vacuum fields. 
In the presence of a cosmological constant the vacuum field equations can be written $R_{\mu\nu}=\Lambda g_{\mu\nu}$. Hence all conclusions drawn from $R_{\mu\nu}k^\mu k^\nu=0$ in section 2 remain valid, we again arrive at the metric (\ref{le}). 
Recently such solutions with $\Lambda \neq 0$ have been found by Firouzhjahi and Mashhoon \cite{Fir}. 
Also non-vacuum gravitational fields with a twist-free null Killing vector have been discussed, e.g. electro-vac solutions \cite{Kram,LPS}; for a survey see \cite{R1}. 


\end{document}